# Designer metal-free altermagnetism in honeycomb two-dimensional frameworks.

Hongde Yu[1], Thomas Brumme[1] and Thomas Heine[1, 2, 3]

1 Faculty of Chemistry and Food Chemistry, Technische Universität Dresden, Bergstraße 66c, 01062 Dresden, Germany.

2 Institute of Resource Ecology, Helmholtz Zentrum Dresden-Rossendorf, Permoserstraße 15, 04318 Leipzig, Germany.

3 Department of Chemistry, Yonsei University, Seodaemun-gu, Seoul 120-749, Republic of Korea.

Email: hongde.yu1@tu-dresden.de; thomas.heine@tu-dresden.de

**Abstract**:

Altermagnetism combines momentum-dependent spin splitting of opposite-spin channels with zero net magnetization, enabling electric-field control of spin transport that is robust against external magnetic fields. Although widely explored in inorganic systems, metal-free altermagnets with π-spin splitting, particularly in two-dimensional (2D) organic frameworks, have remained elusive. Here we introduce a molecular design strategy that achieves designer metal-free altermagnetism in honeycomb 2D crystals. By reducing the monomer point-group symmetry from $D_{3h}$ to $C_{2v}$ in triangulene-derived radicals, inversion symmetry is selectively broken while the bipartite lattice is preserved. Spin-polarized density-functional-theory calculations reveal strong antiferromagnetic couplings of -130 meV, $d$-wave spin splitting of 17 meV at the $M$ point, and Mott-Hubbard gaps of 1.26 eV, all fully consistent with Lieb's theorem. A minimal tight-binding model shows that anisotropic nearest-neighbor hopping arising from direction-dependent π-orbital overlap is the microscopic origin of spin splitting and altermagnetism. Biaxial compressive strain further enhance the spin splitting to 27 meV. These results establish a general approach to room-temperature organic altermagnets and open a pathway toward carbon-based altermagnetism via engineered inversion-symmetry breaking.

**Main text:**

Altermagnetism is an emerging magnetic order that lies between ferromagnetism (FM) and antiferromagnetism (AFM). [1–7] It lifts Kramers spin degeneracy, producing a momentum-dependent spin splitting of opposite-spin channels while preserving zero net magnetization. This combination enables electric-field control of spin transport that remains robust against external magnetic fields.[8] Altermagnets open a rich platform for exploring novel quantum phenomena and realizing ultrafast, field-insensitive spintronic devices. Although altermagnetism has been realized in several inorganic compounds, such as MnTe [1,9], $RuO_2$, [9,10,10] and CrSb [2,11], the design of metal-free altermagnets with π-spin splitting presents a major challenge. [12,13]

Metal-free magnetism relies on the combination of stable paramagnetic centers with precise control of magnetic interactions. [14] Open-shell organic radicals, such as triangulenes (TRI) [15,16] and other polycyclic aromatic hydrocarbons [17–19], serve as versatile building blocks for extended π-conjugated frameworks. [18,20–26] Recent advances in on-surface synthesis have enabled the fabrication of AFM-coupled dimers and 1D spin chains with magnetic interactions reaching -45 meV. [27,28] In TRI-based spin-1 Heisenberg chains, a Haldane gap and fractional edge excitations emerge, [29] whereas spin-1/2 chains assembled from monoradical monomers (phenalenyl, Clar's goblet, and olympicene) exhibit gapless excitations. [17–19,21,22,24,26] Despite these advances in low-dimensional organic magnets,[30–32] an altermagnetic 2D honeycomb lattice analogous to graphene has not yet been proposed.

Here, we propose a molecular strategy to address this challenge by lifting Kramers spin degeneracy in metal-free 2D honeycomb crystals. TRI, possessing a triplet ground state and non-Kekulé character, provides a versatile multiradical platform for engineering metal-free magnetism in honeycomb 2D polymers.[33,34] Hetero-TRI-based 2D polymers have recently been realized experimentally and exhibit many exotic phenomena, including Mott-insulating states, Stoner ferromagnetism, and excitonic insulators.[35–39] However, most TRI and hetero-TRI monomers possess $D_{3h}$ symmetry. Consequently, when these $D_{3h}$ monomers are linked into 2D crystals, the resulting frameworks belong to the high-symmetry space group *P6/mmm*, which intrinsically

possesses an inversion center.[34]

These 2D crystals show half-filled, bipartite lattices and thus obey Lieb's theorem, giving a total spin $S = |N_A - N_B|$.[40] With balanced sublattices $|N_A - N_B|$, the ground state carries zero net magnetization and exhibits AFM coupling between spin centers driven by strong electron correlations. Crucially, the inversion symmetry connecting opposite-spin centers within the unit cell preserves both time-reversal symmetry and Kramers degeneracy, thereby forbidding any momentum-space spin splitting.

In contrast, as shown in Fig. 1, lowering the monomer point-group symmetry to $C_{2v}$ will break the inversion symmetry of these organic 2D crystals. The spin centers remain connected by mirror planes ($m$) and twofold rotation axes ($C_2$), although the spin densities are fully delocalized on the monomers (Fig. 2c). Lieb's theorem nevertheless guarantees AFM coupling between these centers and preserve zero net magnetization. These symmetries precisely fulfill the requirements for altermagnetism, yielding 2D frameworks that belong to the orthorhombic space group *Amm2* without inversion symmetry.

Following this design principle, we performed spin-polarized density-functional theory (DFT) calculations on the altermagnetic 2D crystals [TOT-O] and [TAM-O]. Both molecular building blocks, TOT-O and TAM-O, possess $C_{2v}$ symmetry. (Fig. 2) For every system, the AFM low-spin state ($S = 0$) is the ground state and lies lower in energy than both the FM high-spin ($S = 1$) and diamagnetic closed-shell states, consistent with Lieb's theorem. The spin centers localized on TOT-O and TAM-O exhibit local spin $S = 1/2$, owing to imbalanced sublattices within each monomer.

The magnetic interactions are described by the Heisenberg-Dirac-van Vleck Hamiltonian $\hat{H} = - \sum_{<i, j>} J_{ij} \hat{S}_i \hat{S}_j$, where $J_{ij}$ is the nearest-neighbor coupling constant. As shown in Fig. 2, DFT calculations yield strong AFM couplings of $J = -132$ meV for [TAM-O] and $J = -123$ meV for [TOT-O]. These values are substantially larger than those reported for other organic altermagnets. Within the mean-field approximation, the Néel temperature reaches approximately 2200 K, indicating potential for room-temperature altermagnetism. This remarkably strong AFM coupling originates from the large spatial overlap between the singly occupied molecular orbitals (SOMOs) of neighboring spin centers, enabled by the extended π-conjugated character of the organic radicals. The considerable electronic hopping integral $t \approx 0.2$ eV leads to a dominating kinetic-exchange of $-4t^2 \approx -0.16$ eV, fully consistent with the large $J$ extracted from first-principles calculations.

The spin-polarized band structures of the altermagnetic 2D crystals [TOT-O] and [TAM-O] confirm that Kramers spin degeneracy is lifted, consistent with the broken inversion symmetry (Fig. 2). At the $M$ (0, 0.5) point, the spin-splitting reaches 14 meV for [TOT-O] and 17 meV for [TAM-O], while the spin-up and spin-down channels remain degenerate at the $\varGamma$ and $K$ points. Both systems exhibit a large insulating gap of approximately 1.26 eV, originating from on-site Coulomb repulsion of the strongly correlated π-electrons. In contrast, the diamagnetic closed-shell band structure is nearly gapless (approximately 0.06 eV), underscoring the Mott-Hubbard insulating character[23,41] induced by AFM ordering (Fig. 3).

At the $M$ point, both the valence band maximum (VBM) and conduction band minimum (CBM) belong to the spin-down channel. Consequently, momentum-selective optical excitation can generate a pure spin current via spin-conserving transitions, different from metallic altermagnets. As shown in Fig. 2, the resulting spin-polarized band structure, especially the spin splitting at $M$ (0, 0.5) and $M'$ (0.5, 0) exhibits the characteristic $d$-wave symmetry of altermagnetism. Extending the linkage from direct bonding to –CC–, –CCCC– spacers systematically reduces both the magnetic coupling $J$ (-86 and -68 meV) and the spin-splitting (8 and 6 meV) magnitude.

The microscopic origin of metal-free altermagnetism in these 2D crystals lies in anisotropic nearest-neighbor hopping (Fig. 3). Employment of $C_{2v}$-symmetric spin units breaks the inversion symmetry of the honeycomb lattice, resulting in direction-dependent hopping amplitudes. This anisotropy arises from the unequal overlap of the $p_z$-components of the SOMOs along different directions. In the closed-shell band structure, a band gap opens, driven by the this anisotropic hopping together with the breaking of $C_3$ rotational symmetry. A minimal tight-binding (TB) Hamiltonian with unequal nearest-neighbor hopping parameters ($t_1 = 0.19$ eV, $t_2 = 0.23$ eV) preproduces the DFT band structures (Fig. 3b). The resulting asymmetric dispersion and bandgap opening confirm that anisotropic hopping is the essential mechanism underlying altermagnetism in these systems. This insight suggests a viable route toward carbon-based altermagnets. Inversion-symmetry breaking via vacancies, periodic superlattices, or Moiré patterns can similarly induce anisotropic hopping in graphene-derived systems.

External strain provides an additional handle to modulate the spin splitting in these altermagnetic 2D crystals. Biaxial strain applied along both lattice vectors $a$ and $b$ modulate the spin-splitting energy at the $M$ point. As shown in Fig. 3, the spin splitting energies $\Delta E_{SS}$ varies linearly with strain. For [TAM-O], compressive strain of -5% increases the splitting from 17 meV to 27 meV while simultaneously strengthening the AFM coupling to $J = -153$ meV. In contrast, tensile strain reduces the spin splitting energies because the staggered potential is partially averaged out, diminishing the contribution of anisotropic hopping.

**Conclusion:**

In summary, we have introduced a molecular design strategy for designer metal-free altermagnetism in honeycomb 2D organic frameworks. By lowering the monomer symmetry from $D_{3h}$ to $C_{2v}$ while preserving a bipartite lattice, the resulting [TOT-O] and [TAM-O] crystals break inversion symmetry and lift Kramers spin degeneracy, yielding $d$-wave spin splitting of 14-17 meV at the $M$ point, strong AFM coupling ($J \approx -130$ meV), and Mott-Hubbard insulating gaps of approximately 1.26 eV, all fully consistent with Lieb's theorem and a minimal tight-binding model of anisotropic hopping. Biaxial compressive strain further enhances the spin splitting energies to 27 meV and $J$ to -153 meV, demonstrating robust tunability. These findings establish a general route toward room-temperature, electrically controllable spintronic devices based on purely organic altermagnets and open a pathway to graphene-derived altermagnetism via engineered inversion-symmetry breaking.

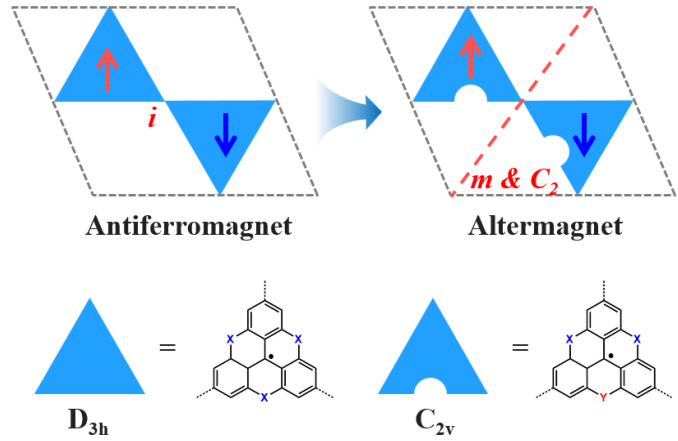

**Fig. 1.** Molecular design strategy for designer metal-free altermagnetism in honeycomb 2D organic frameworks.

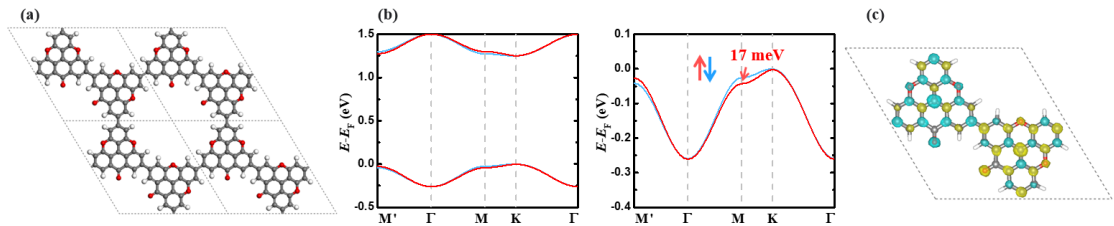

**Fig. 2.** (a) Crystal structure, (b) spin-polarized band structure of the ground state, and (c) spin density distribution of the metal-free altermagnetic 2D frameworks [TAM-O]. $M$: (0, 0.5); $M'$: (0.5, 0).

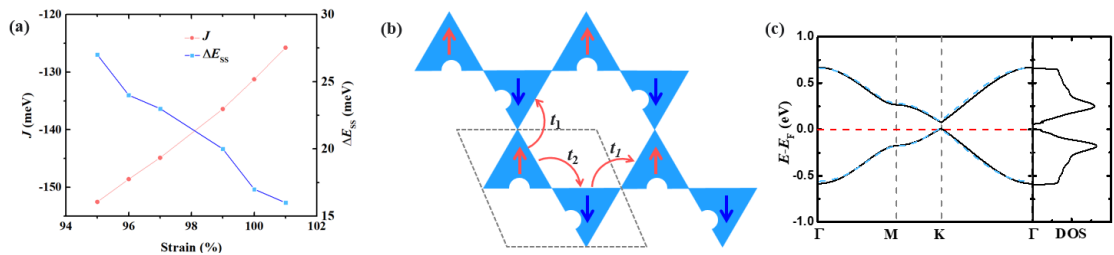

**Fig. 3.** Mechanism and modulation of metal-free altermagnetism 2D honeycomb lattice. (a) Magnetic coupling $J$ and spin-splitting energy $\Delta E_{\mathrm{SS}}$ as a function of biaxial strain for [TAM-O]. (b) Schematic illustration of the microscopic origin of altermagnetism arising from anisotropic nearest-neighbor hopping of minimal tight-binding model in the $C_{2v}$-symmetric honeycomb lattice. (c) Band structure calculated from DFT and the tight-binding Hamiltonian with two distinct hopping parameters ($t_1 = 0.19$ eV, $t_2 = 0.23$ eV) for [TAM-O].